\documentclass{epl} 
\usepackage[dvips]{epsfig}
\begin{document}

\title{Lattice Boltzmann simulations of apparent slip in hydrophobic microchannels}
\shorttitle{LB simulations of apparent slip}
\author{Jens Harting, Christian Kunert, and Hans J. Herrmann}
\institute{Institute for Computational Physics, University of Stuttgart, Pfaffenwaldring 27, D-70569
Stuttgart, Germany}
\date{\today}
\maketitle

\begin{abstract}
Various experiments have found a boundary slip in hydrophobic microchannel
flows, but a consistent understanding of the results is still lacking.
While Molecular Dynamics (MD) simulations cannot reach the low
shear rates and large system sizes of the experiments, it is often
impossible to resolve the needed details with macroscopic approaches. We
model the interaction between hydrophobic channel walls and a fluid by
means of a multi-phase lattice Boltzmann model. Our mesoscopic approach
overcomes the limitations of MD simulations and can reach the small flow
velocities of known experiments. We reproduce results from experiments at
small Knudsen numbers and other simulations, namely an increase
of slip with increasing liquid-solid interactions, the slip being
independent of the flow velocity, and a decreasing slip with increasing
bulk pressure. Within our model we develop a semi-analytic approximation
of the dependence of the slip on the pressure.
\end{abstract}

During the last century it was widely assumed that the velocity of a
Newtonian liquid at a surface is always identical to the velocity of the
surface. However, in recent years well controlled experiments have
shown a violation of the no-slip boundary condition in sub-micron sized
geometries. Since then, experimental
\cite{bib:lauga-brenner-stone} and
theoretical works
\cite{bib:theory}, 
as well as computer simulations
~\cite{bib:lbsims,bib:barrat-bocquet-99,bib:koplik-banavar-willemsen-89,koplik-banavar-98,bib:cieplak-koplik-banavar-01,bib:thompson-robbins-1990,bib:troian97}
have tried to improve our understanding of boundary slip.
The complex behavior of a fluid close to a solid
interface involves the interplay of many physical and chemical properties.
These include the wettability of the solid, shear rate, pressure,
 surface charge, surface roughness, as well as impurities and
dissolved gas. Since all those quantities have to be determined very
precisely, it is not surprising that our understanding of the phenomenon
is still unsatisfactory. Due to the large number of different
parameters, a significant dispersion of the
results can be observed for ostensibly similar
systems~\cite{bib:lauga-brenner-stone}, e.g. observed slip
lengths vary between nanometres~\cite{bib:churaev-sobolev-somov-84}
and micrometers
~\cite{bib:tretheway-meinhart-0204} 
and while some authors find a dependence of the slip on the flow
velocity
~\cite{bib:varvel}, 
others do not
~\cite{cheng-giordano-02,bib:tretheway-meinhart-0204}.
Most computer simulations apply Molecular Dynamics (MD) 
and report increasing slip with decreasing liquid density
\cite{bib:koplik-banavar-willemsen-89,bib:thompson-robbins-1990} or
liquid-solid interactions
\cite{bib:cieplak-koplik-banavar-01,bib:nagayama-cheng-2004},
while slip decreases with increasing pressure
\cite{bib:barrat-bocquet-99}. These simulations are usually limited to
some tens of thousands of particles, lengths scales of nanometres and
timescales of nanoseconds. Also, shear rates are orders of
magnitude higher than in any experiment~\cite{bib:lauga-brenner-stone}.
We overcome these limitations using
the lattice Boltzmann (LB) algorithm -- a powerful method for
simulating fluid
dynamics~\cite{bib:succi}.
Rather than tracking
individual atoms and molecules, the
dynamics of the single-particle distribution function $\eta$ of mesoscopic
fluid packets is described. In contrast to MD simulations, this method is
less computationally demanding and allows to simulate
experimentally accessible length and time scales.
Our ansatz differs from other LB approaches
where slip is introduced by generalizing no-slip bounce back
boundary conditions to allow specular reflections with a given
probability
~\cite{bib:lbsims} or where the viscosity is modified due to local
density variations~\cite{bib:nie-doolen-chen}. In both cases, parameters
determining the properties at the boundaries are not easily mappable to
experimentally available values. 
Our approach is based on Shan and Chen's multi-phase LB model
~\cite{bib:shan-chen}.
Here, interactions
between different species are modelled by mesoscopic forces between
the phases. This naturally opens the way to introduce similar
interactions between each fluid species and the channel walls, where the
strength of the interaction is determined by the fluid densities, free
coupling constants, and a wall interaction parameter which is treated in a
similar manner as a local fluid density. 
The model allows the simulation of multi-phase flows along hydrophobic
boundaries and is introduced in the following. However, in
order to study the influence of hydrophobicity on the boundary slip and to
demonstrate the basic properties of the model, we
focus on single phase flow in this paper. Results of multi-phase
simulations will be presented in a future work. 
A multi-phase LB system can be represented by a set of
equations
~\cite{bib:mplb}
\begin{equation}
\label{LBeqs}
\begin{array}{cc} 
\eta_i^{\alpha}({\bf x}+{\bf c}_i, t+1) - \eta_i^{\alpha}({\bf x},t) =
\Omega_i^{\alpha}, &  i= 0,1,\dots,b\mbox{ ,} 
\end{array}
\end{equation}
where $\eta_i^{\alpha}({\bf x},t)$ is the single-particle distribution
function, indicating the amount of species $\alpha$ with velocity ${\bf c}_i$, at site ${\bf x}$
on a D-dimensional lattice of coordination number $b$ (D3Q19 in our
implementation), at time-step
$t$. For the collision operator $\Omega_i^{\alpha}$ we choose the 
Bhatnagar-Gross-Krook (BGK) form
\begin{equation}
\label{Omega}
 \Omega_i^{\alpha} =
 -\frac{1}{\tau^{\alpha}}(\eta_i^{\alpha}({\bf x},t) - \eta_i^{\alpha
\, eq}({\bf u}^{\alpha}({\bf x},t),\eta^{\alpha}({\bf x},t)))\mbox{ ,}
\end{equation}
where $\tau^{\alpha}$ is the mean collision time for component $\alpha$
and determines the fluid viscosity.
The system relaxes to an equilibrium distribution $\eta_i^{\alpha\,eq}$
which can be derived imposing restriction on
the microscopic processes, such as explicit mass and momentum
conservation for each
species
~\cite{bib:lbneq}.
$\eta^{\alpha}({\bf x},t)\equiv\sum_i \eta_i^{\alpha}({\bf x},t)$ is the
fluid density and 
${\bf u}^{\alpha}({\bf x},t)$ is 
the macroscopic velocity of the fluid, defined as
$\eta^{\alpha}({\bf x},t){\bf u}^{\alpha}({\bf x},t)
\equiv \sum_i \eta_i^{\alpha}({\bf x},t){\bf c}_i$.
%
Interactions between different fluid species are introduced as a
mean field body force between nearest
neighbors
~\cite{bib:shan-chen}:
\begin{equation}
\label{Eq:force}
{\bf F}^{\alpha \bar{\alpha}}({\bf x},t) \equiv -\psi^{\alpha}({\bf x},t)\sum_{\bf
\bar{\alpha}}g_{\alpha \bar{\alpha}}\sum_{\bf
x^{\prime}}\psi^{\bar{\alpha}}({\bf x^{\prime}},t)({\bf x^{\prime}}-{\bf
x})\mbox{ ,}
\end{equation}
where $\psi^{\alpha}({\bf x},t)=(1 - e^{-\eta^{\alpha}({\bf
x},t)/\eta_0})$ is the
so-called effective mass with $\eta_0$ being
a reference density that is set to 1 in our case~\cite{bib:shan-chen}.
$g_{\bar{\alpha}\alpha}$ is a force coupling constant, whose magnitude
controls the strength of the interaction between component $\alpha$
and $\bar{\alpha}$. The
dynamical effect of the force is realized in the BGK collision
operator in Eq.~(\ref{Omega}) by adding to the velocity ${\bf u}$ in the
equilibrium distribution 
an increment
$\delta{\bf u}^{\alpha} = {\tau^{\alpha}{\bf
F}^{\alpha \bar{\alpha}}}/{\eta^{\alpha}}$.
For the interaction of the fluid components with the channel walls we
apply mid-grid bounce back boundary conditions~\cite{bib:succi} and assign
interaction properties to the wall which are similar to those of an
additional fluid species. I.e., we specify constant values for the force
coupling constant $g_{\bar{\alpha}\alpha}=g_{wall,\alpha}$ and the density
$\eta^{\bar{\alpha}}=\eta^{wall}$ at wall boundary nodes of the lattice. This
results in a purely local force as given in Eq.~\ref{Eq:force} between the
flow and the boundaries. Even though one could argue that a single
parameter to tune the fluid-wall interaction would be sufficient, we keep
our approach as
close as possible to the original idea of Shan and Chen in order to
benefit from the experience obtained from other works using the original
model. Furthermore, the additional parameter allows more flexibility when
simulating not only a single fluid, but a multi-phase system.
The fluid-wall interaction can be linked to a contact angle between fluid
droplets and solid walls as it is
often used to quantitatively describe hydrophobic
interactions
~\cite{bib:contact-angle}. Recently, Benzi et al. have shown how to
compute the contact angle within the Shan-Chen model~\cite{bib:benzi-06}. 
The same authors also developed an approach to model apparent slip which
is related to ours, but instead of using only local fluid-solid interactions, they
add an exponential decay of the interaction with distance from the
wall~\cite{benzi-biferale-05}.
We simulate pressure driven flow between two infinite planes (Poiseuille
flow), where pressure driven boundary conditions are implemented in a
similar way as in most experiments: a fixed pressure is set at the channel
inlet and an open boundary at the outlet. The outlet is realized by
interpolating the particle distribution function at the end of the channel
as given by $ \eta_i^{\alpha}({\bf x},t)=2\eta_i^{\alpha}({\bf
x-1},t)-\eta_i^{\alpha}({\bf x-2},t)$ leading to a linear pressure
gradient.
Already in 1823, Navier proposed a boundary condition where the fluid
velocity at a surface is proportional to the shear rate at the
surface, i.e. $v_z(x_0)=\beta\partial v_z(x)/\partial x$ at $x=x_0$~\cite{bib:Navier}. 
Following his hypothesis the velocity in flow
direction ($v_z$) at position $x$ between the planes is given by
\begin{equation} 
v_z(x)=\frac{1}{2 \mu}\frac{\partial P}{\partial z}
\left[ h^2-x^2-2h\beta  \right],
\label{eq:plattenprofil} 
\end{equation} 
where $2h$ is the distance between the planes, and $\mu$ the viscosity. In
contrast to a no-slip formulation, the last term in
Eq.~\ref{eq:plattenprofil} linearly depends on the slip length $\beta$.
Since $\beta$ is typically of the order of nanometers or micrometers, it
can be neglected in macroscopic experiments. 
In order to obtain $\beta$ from our data, we measure the pressure
gradient ${\partial P}/{\partial z}$ at the center of the channel and 
the velocity profile between the two planes at a fixed position $z$.
$\beta$ is then obtained by a least square fit with Eq.~\ref{eq:plattenprofil}.
 
Our simulation parameters are as follows: the lattice size is kept
constant with the channel length ($z$ direction) being 256 sites, the
distance between the plates $2h$ being 60 sites ($x$ direction). We
approximate infinite planes by using a 16 sites wide channel with periodic
boundaries in $y$ direction. In order to assure a fully
equilibrated system we simulate for at least 40000 time steps before
measuring and assured our results being independent of the discretization
level by comparing to simulations of 28 and 124 sites wide channels.
Each data point in the figures below corresponds to about six
hours simulation time on eight IBM Power 4 1.7GHz CPUs. All units in this
paper are in lattice units with the lattice constant $c$ and timestep
$\Delta t$ set to 1 if not stated otherwise.
\begin{figure}
\centerline{\epsfig{file=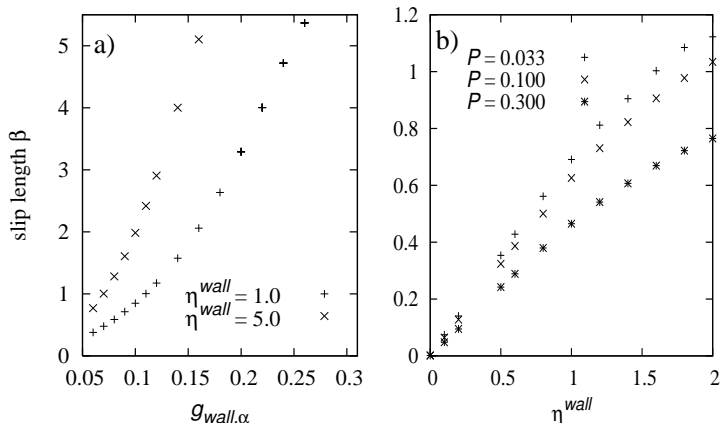,angle=270,width=0.70\linewidth}}
\caption{\label{fig:slipvsgbr} Slip versus $g_{wall,\alpha}$ for different
wall interactions $\eta^{wall}$ and constant $P$=0.11, $V$=0.033 (a).
$\beta$ is steadily increasing with increasing $g_{wall,\alpha}$ and
achievable slip lengths are higher for a larger $\eta^{wall}$. Fig.~b) 
shows $\beta$ versus interaction parameter
$\eta^{wall}$ for various bulk pressures and fixed $V=3.5\cdot 10^{-3}$.
For lower pressure, larger values of $\beta$ are measured.\label{fig:betavsRC}}
\end{figure}

The dependence of the slip length $\beta$ on the interaction parameter
$g_{wall,\alpha}$ is studied for $\eta^{wall}$=1.0 and 5.0. The bulk
pressure $P=\rho c_s^2$, where $\rho$ is the fluid density and
$c_s=1/\sqrt{3}$ the speed of sound, is kept at
$P$=0.11, while the flow velocity is set to $V$=0.033. As shown in
Fig.~\ref{fig:slipvsgbr}a we vary $g_{wall,\alpha}$ from 0.06 to 0.22
and find a steady increase of $\beta$ for increasing $g_{wall,\alpha}$. As
expected, the curve for $\eta^{wall}$=5.0 is growing substantially faster
than for $\eta^{wall}$=1.0. The maximum available $\beta$ are at about 5.2
for $g_{wall,\alpha}$=0.26 and $\eta^{wall}$=1.0. At these strong
fluid-wall 
interactions, the force as given in Eq.~\ref{Eq:force}
becomes very large and results in a large area of low fluid density close
to the wall. Increasing the interaction even further results in numerical
instabilities due to too steep density gradients. In order to study the dependence of the slip on other
parameters, the coupling constant $g_{wall,\alpha}$ is kept constant at
0.08 from now on.
Fig.\ref{fig:slipvsgbr}b depicts the dependence of $\beta$ on $\eta^{wall}$
for different bulk pressures $P$=0.033, 0.1, and 0.3 and fixed flow
velocity $V=3.5\cdot 10^{-3}$ in the system. While all three graphs grow
constantly with increasing $\eta^{wall}$, the one for $P$=0.033 grows the
fastest demonstrating that absolute values for $\beta$ are
higher for lower pressure.

\begin{figure}[h]
\centerline{\epsfig{file=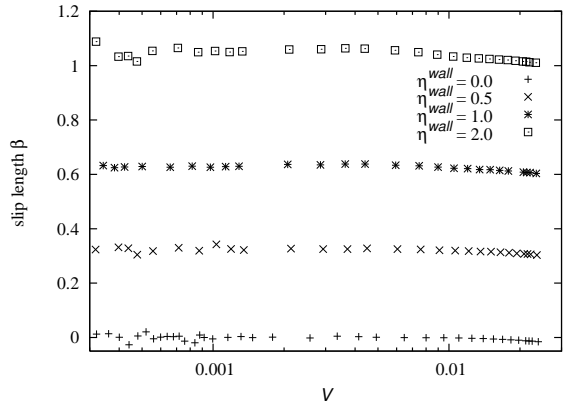,angle=270,width=0.55\linewidth}}
\caption{\label{fig:betavsv} Slip length $\beta$ versus flow velocity $V$
for different wall interactions $\eta^{wall}$.
While we do not find any slip for $\eta^{wall}=$0.0, $\beta$ increases
with increasing $\eta^{wall}$. We vary the flow velocity from $1\cdot
10^{-4}$ to 0.03 and find constant values for $\beta$ independent of 
$V$ (within numerical accuracy).}
\end{figure}
We have measured the magnitude of the boundary slip over a very wide range
of flow velocities $V$ from $1\cdot 10^{-4}$ to $3\cdot 10^{-2}$ for wall
interactions $\eta^{wall}$=0.0, 0.5, 1.0, and 2.0. $V$ is measured at the
center of the channel and given on a logarithmic scale in
Fig.~\ref{fig:betavsv}. For $\eta^{wall}$=0.0 we do not find any boundary
slip confirming that our method properly reproduces no slip behavior in
the interaction free case. With increasing wall interactions, we achieve
an increase of the magnitude of $\beta$ to up to $\simeq$1.1 for
$\eta^{wall}$=2.0. We are not able to find any velocity dependence of
$\beta$, but find constant slip for fixed fluid-wall interactions, which is consistent with many
experiments~\cite{cheng-giordano-02,cheikh-koper-03}. The fluctuations of the data for very low flow velocities are
due to numerical uncertainties of the fit at very low curvature
of the parabolic velocity profile. For $V>0.01$ we find a slight
deviation of $\beta$ from the constant measurements. This is due to a
small variation of the bulk pressure from $P$=0.097 for $V=1\cdot 10^{-4}$
to $P$=0.106 for $V=0.03$ that cannot easily be avoided for technical reasons.
We have checked for a few data points that $\beta$ stays constant if 
$P$ can be kept at exactly fixed values, too.
The slip length being independent of the flow velocity is consistent with
many experiments and computer simulations, like the MD simulations of
Cottin-Bizonne et
al.
~\cite{bib:cottin-bizone} 
and the experiments of Cheng et al.~\cite{cheng-giordano-02} and Baudry et
al.~\cite{baudry-charlaix-01}. We speculate that an increase of $\beta$
with increasing flow velocity as measured by some experiments
~\cite{bib:varvel} 
is due to surface
roughness of the channel boundaries or other nonlinear effects. Since our
model is not able to treat roughness on an atomic scale, we do not expect
to conform with those results. MD simulations which find a non-constant
value for $\beta$ operate at very high shear rates which are orders of
magnitude higher than what can be obtained by our
approach~\cite{bib:troian97}.

\begin{figure}
\centerline{\epsfig{file=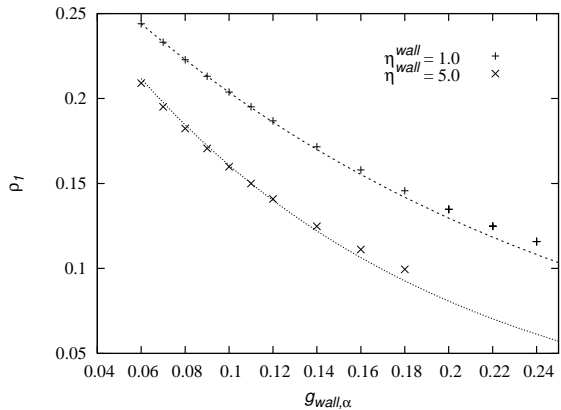,angle=270,width=0.55\linewidth}}
\caption{\label{fig:rhowall}The fluid density close to the channel walls
$\rho_1$ over $g_{wall,\alpha}$ is given for $\eta^{wall}$=1.0, 5.0
(symbols). The lines correspond to a fit by our semi-analytic
approximation.}
\end{figure}
Computing the exact slip in dependence of the interaction parameters from
first principle analytically is a very hard or even impossible task since
our interaction as given in Eq.~\ref{Eq:force} modifies the equilibrium
distribution in the BGK operator. Therefore, we present a semi-analytic
approximation which utilizes the common two-layer model. 
Here, it is assumed that a thin fluid layer with thickness $\delta$ and
different viscosity as the bulk fluid exists near the channel walls. As
calculated by various authors
~\cite{bib:theory}, 
within this model the slip length can be computed as
$\beta=(\mu_{bulk}/\mu_1-1)\delta$, where $\mu_{bulk}$ is the viscosity of
the bulk fluid, and $\mu_1$ the viscosity close to the wall.
Since the dynamic viscosity is given by
the kinematic viscosity times the fluid density,
$\mu=\rho\nu=\rho(2\tau^\alpha -1)/6$~\cite{bib:succi},
we write $\beta=(\rho_{bulk}/\rho_1-1)\delta$.
$\rho_{bulk}$ can be measured in the channel's center
and $\rho_1$ at the first lattice site next to the wall.
Fig.~\ref{fig:rhowall} shows the dependence of $\rho_1$ on
$g_{wall,\alpha}$ for $\eta^{wall}$=1.0, 5.0, $P$=0.11, and $V$=0.033.
Since $\rho_1$ cannot
easily be computed analytically, we
postulate an interaction term that depends on the bulk density and the
fluid-wall interaction as well as a free fit parameter $k$,
\begin{equation}
{\cal I}=k{\bf F}^{wall,\alpha}({\bf x},t)/\rho_{bulk}({\bf x},t)
\end{equation}
and fit $\rho_1$ with an exponential function
$\rho_1=\rho_{bulk}({\bf x},t)\exp(-{\cal I})$.
With only a single value for $k$ we are able to utilize
this equation
to fit $\rho_1$ for all our simulation parameters. $k$ is found to be 8.35
for our data. The lines in Fig.~\ref{fig:rhowall} illustrate the good
quality of our approximation. A similar approach is applied to model the
thickness of the layer at the wall which strongly depends on the
fluid-wall interaction and bulk density. Here, we set $\delta=\exp({\cal
I})$. As a result, $\beta$ can be estimated by 
$\beta=(\exp({\cal I})-1)\exp({\cal I})$.
\begin{figure}
\centerline{\epsfig{file=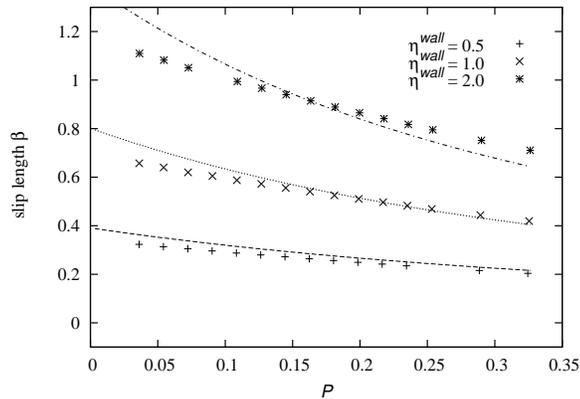,angle=270,width=0.55\linewidth}}
\caption{\label{fig:betavsp} Slip length $\beta$ versus bulk pressure $P$
for $\eta^{wall}=0.5, 1.0, 2.0$ (symbols). $\beta$ 
increases with increasing fluid-wall interactions, but decreases with
increasing $P$. This dependence
can be described by a semi-analytic equation (lines) which agrees well for
small fluid-wall interactions and qualitatively reproduces our
data for strong fluid-wall interactions.}
\end{figure}
The semi-analytic approximation is used to fit 
the dependence of the slip length $\beta$ on
the bulk pressure $P$. Fig.~\ref{fig:betavsp} shows the simulation data
(symbols) and the approximation (lines) for wall interactions
$\eta^{wall}$=0.5, 1.0, and 2.0. The bulk pressure is varied from 0.03 to
0.33. We find a
decrease of $\beta$ with increasing pressure $P$. An increase of
$\eta^{wall}$ leads to an increasing slope of the curves and to higher
absolute values for $\beta$. Furthermore, we find a decrease
of the slip
with increasing bulk pressure. These results qualitatively agree with MD
simulations~\cite{bib:barrat-bocquet-99,bib:cieplak-koplik-banavar-01}.
Even with a single value for the fit parameter $k$, the semi-analytic
description of $\beta$ agrees very well for low fluid-wall interactions.
For strong interactions ($\eta^{wall}$=2.0), the fit qualitatively
reproduces the behavior of the slip length. Higher order terms in the
exponential ansatz for $\delta$ are needed for a better agreement.

To demonstrate that our approach is able to achieve experimentally
available length and time scales, we scale our simulations to the
experimental setup of Tretheway and Meinhart
~\cite{bib:tretheway-meinhart-0204}. 
They use a 30$\mu$m high and 300$\mu$m wide microchannel with typical flow
velocities of $V=10^{-2}$mm/s. For water, they measure a slip length of
0.92$\mu$m. The Reynolds number $Re=2hV/\nu$ in their experiment is
$\simeq$0.3. To reproduce the observed slip, we set
$g_{wall,\alpha}$=0.16 and $\eta^{wall}=1.0$ (see
Fig.~\ref{fig:slipvsgbr}a).
We are able to cover a wide range of flow velocities,
i.e. for the setup given above, velocities can range from as low as
$1\times 10^{-4}$ to as high as 0.05 corresponding to $Re$ 
between 0.038 and 19. 
The Knudsen number is given by $Kn=\nu/(c_s 2h)$ which corresponds
to $4.8\times 10^{-3}$ for the simulations presented here. At these low
$Kn$, the hydrodynamic approach is well valid. However, it has been 
shown that the LB method can be applied for
$Kn$ much larger than 1 if one uses modified boundary
conditions~\cite{toschi-succi-05}. Our mesoscopic force is expected to be
able to properly describe fluid-wall interactions in such systems as well. 

In conclusion, we presented a new approach to investigate boundary
slip in hydrophobic microchannels by means of a multi-phase LB model. In
contrast to MD simulations, our model is able to reach the length and time
scales of typical experiments and is applicable for a wide range of
realistic flow velocities. We qualitatively reproduced the dependence
of slip on the hydrophobicity of the channel walls and
found constant slip for varying flow velocities. The decrease of the
slip with increasing pressure can be approximated by a semi-analytic
approach. Our results are consistent with MD simulations
~\cite{bib:cottin-bizone,bib:barrat-bocquet-99,bib:cieplak-koplik-banavar-01}
and experiments
~\cite{bib:tretheway-meinhart-0204}.
\begin{acknowledgments}
We would like to thank G.~Giupponi, M.~Hecht, N.~Gonz\'alez-Segredo, and
V.S.J.~Craig for fruitful discussions and acknowledge the Neumann
Institute for Computing for providing access to their IBM p690 system.
\end{acknowledgments}

\begin{thebibliography}{40}
\expandafter\ifx\csname natexlab\endcsname\relax\def\natexlab#1{#1}\fi
\expandafter\ifx\csname bibnamefont\endcsname\relax
  \def\bibnamefont#1{#1}\fi
\expandafter\ifx\csname bibfnamefont\endcsname\relax
  \def\bibfnamefont#1{#1}\fi
\expandafter\ifx\csname citenamefont\endcsname\relax
  \def\citenamefont#1{#1}\fi
\expandafter\ifx\csname url\endcsname\relax
  \def\url#1{\texttt{#1}}\fi
\expandafter\ifx\csname urlprefix\endcsname\relax\def\urlprefix{URL }\fi
\providecommand{\bibinfo}[2]{#2}
\providecommand{\eprint}[2][]{\url{#2}}

\bibitem[1]{bib:lauga-brenner-stone}
\bibinfo{author}{\bibfnamefont{E.}~\bibnamefont{Lauga}},
  \bibinfo{author}{\bibfnamefont{M.~P.} \bibnamefont{Brenner}},
  \bibnamefont{and} \bibinfo{author}{\bibfnamefont{H.~A.} \bibnamefont{Stone}},
  \emph{\bibinfo{title}{in Handbook of Experimental Fluid Dynamics, edited by
  J. Foss, C. Tropea and A. Yarin}} (\bibinfo{publisher}{Springer},
  \bibinfo{year}{2005}), chap.~\bibinfo{chapter}{15}.

\bibitem[2]{bib:theory}
\bibinfo{author}{\bibfnamefont{P.~G.} \bibnamefont{{De Gennes}}},
  \bibinfo{journal}{Langmuir} \textbf{\bibinfo{volume}{18}},
  \bibinfo{pages}{3413} (\bibinfo{year}{2002}).
%
\bibinfo{author}{\bibfnamefont{O.~I.} \bibnamefont{Vinogradova}},
  \bibinfo{journal}{Langmuir} \textbf{\bibinfo{volume}{11}},
  \bibinfo{pages}{2213} (\bibinfo{year}{1995}).

\bibitem[3]{bib:lbsims}
\bibinfo{author}{\bibfnamefont{S.}~\bibnamefont{Succi}},
  \bibinfo{journal}{Phys. Rev. Lett.} \textbf{\bibinfo{volume}{89}},
  \bibinfo{pages}{064502} (\bibinfo{year}{2002}).
%
%
\bibinfo{author}{\bibfnamefont{D.~C.} \bibnamefont{Tretheway}},
  \bibinfo{author}{\bibfnamefont{L.}~\bibnamefont{Zhu}},
  \bibinfo{author}{\bibfnamefont{L.}~\bibnamefont{Petzold}}, \bibnamefont{and}
  \bibinfo{author}{\bibfnamefont{C.~D.} \bibnamefont{Meinhart}}, in
  \emph{\bibinfo{booktitle}{Proc. of IMECE}} (\bibinfo{year}{2002}).

\bibitem[4]{bib:barrat-bocquet-99}
\bibinfo{author}{\bibfnamefont{J.-L.} \bibnamefont{Barrat}} \bibnamefont{and}
  \bibinfo{author}{\bibfnamefont{L.}~\bibnamefont{Bocquet}},
  \bibinfo{journal}{Phys. Rev. Lett.} \textbf{\bibinfo{volume}{82}},
  \bibinfo{pages}{4671} (\bibinfo{year}{1999}).

\bibitem[5]{koplik-banavar-98}
\bibinfo{author}{\bibfnamefont{J.}~\bibnamefont{Koplik}} \bibnamefont{and}
  \bibinfo{author}{\bibfnamefont{J.~R.} \bibnamefont{Banavar}},
  \bibinfo{journal}{Phys. Rev. Lett.} \textbf{\bibinfo{volume}{80}},
  \bibinfo{pages}{5125} (\bibinfo{year}{1998}).

\bibitem[6]{bib:koplik-banavar-willemsen-89}
\bibinfo{author}{\bibfnamefont{J.}~\bibnamefont{Koplik}},
  \bibinfo{author}{\bibfnamefont{J.~R.} \bibnamefont{Banavar}},
  \bibnamefont{and} \bibinfo{author}{\bibfnamefont{J.~F.}
  \bibnamefont{Willemsen}}, \bibinfo{journal}{Phys. Fluids}
  \textbf{\bibinfo{volume}{1}}, \bibinfo{pages}{781} (\bibinfo{year}{1989}).

\bibitem[7]{bib:thompson-robbins-1990}
\bibinfo{author}{\bibfnamefont{P.~A.} \bibnamefont{Thompson}} \bibnamefont{and}
  \bibinfo{author}{\bibfnamefont{M.~O.} \bibnamefont{Robbins}},
  \bibinfo{journal}{Phys. Rev. A} \textbf{\bibinfo{volume}{41}},
  \bibinfo{pages}{6830} (\bibinfo{year}{1990}).

%
\bibitem[8]{bib:cieplak-koplik-banavar-01}
\bibinfo{author}{\bibfnamefont{M.}~\bibnamefont{Cieplak}},
  \bibinfo{author}{\bibfnamefont{J.}~\bibnamefont{Koplik}}, \bibnamefont{and}
  \bibinfo{author}{\bibfnamefont{J.~R.} \bibnamefont{Banavar}},
  \bibinfo{journal}{Phys. Rev. Lett.} \textbf{\bibinfo{volume}{86}},
  \bibinfo{pages}{803} (\bibinfo{year}{2001}).

\bibitem[9]{bib:troian97}
\bibinfo{author}{\bibfnamefont{P.~A.}~\bibnamefont{Thompson}} \bibnamefont{and}
  \bibinfo{author}{\bibfnamefont{S.}~\bibnamefont{Troian}},
  \bibinfo{journal}{Nature} \textbf{\bibinfo{volume}{389}},
  \bibinfo{pages}{360} (\bibinfo{year}{1997}).

\bibitem[10]{bib:churaev-sobolev-somov-84}
\bibinfo{author}{\bibfnamefont{N.~V.} \bibnamefont{Churaev}},
  \bibinfo{author}{\bibfnamefont{V.~D.} \bibnamefont{Sobolev}},
  \bibnamefont{and} \bibinfo{author}{\bibfnamefont{A.~N.} \bibnamefont{Somov}},
  \bibinfo{journal}{J. Colloid Interface Sci.} \textbf{\bibinfo{volume}{97}},
  \bibinfo{pages}{574} (\bibinfo{year}{1984}).

\bibitem[11]{bib:tretheway-meinhart-0204}
\bibinfo{author}{\bibfnamefont{D.~C.} \bibnamefont{Tretheway}}
  \bibnamefont{and} \bibinfo{author}{\bibfnamefont{C.~D.}
  \bibnamefont{Meinhart}}, \bibinfo{journal}{Phys. Fluids}
  \textbf{\bibinfo{volume}{14}}, \bibinfo{pages}{L9} (\bibinfo{year}{2002}).
%
\bibinfo{author}{\bibfnamefont{D.~C.} \bibnamefont{Tretheway}}
  \bibnamefont{and} \bibinfo{author}{\bibfnamefont{C.~D.}
  \bibnamefont{Meinhart}}, \bibinfo{journal}{Phys. Fluids}
  \textbf{\bibinfo{volume}{16}}, \bibinfo{pages}{1509} (\bibinfo{year}{2004}).

\bibitem[12]{bib:varvel}
\bibinfo{author}{\bibfnamefont{Y.}~\bibnamefont{Zhu}} \bibnamefont{and}
  \bibinfo{author}{\bibfnamefont{S.}~\bibnamefont{Granick}},
  \bibinfo{journal}{Phys. Rev. Lett.} \textbf{\bibinfo{volume}{88}},
  \bibinfo{pages}{106102} (\bibinfo{year}{2002}).

%
\bibinfo{author}{\bibfnamefont{C.~H.} \bibnamefont{Choi}},
  \bibinfo{author}{\bibfnamefont{K.~J.} \bibnamefont{Westin}},
  \bibnamefont{and} \bibinfo{author}{\bibfnamefont{K.~S.}
  \bibnamefont{Breuer}}, \bibinfo{journal}{Phys. Fluids}
  \textbf{\bibinfo{volume}{15}}, \bibinfo{pages}{2897} (\bibinfo{year}{2003}).
%
\bibinfo{author}{\bibfnamefont{V.~S.~J.} \bibnamefont{Craig}},
  \bibinfo{author}{\bibfnamefont{C.}~\bibnamefont{Neto}}, \bibnamefont{and}
  \bibinfo{author}{\bibfnamefont{D.~R.~M.} \bibnamefont{Williams}},
  \bibinfo{journal}{Phys. Rev. Lett.} \textbf{\bibinfo{volume}{87}},
  \bibinfo{pages}{054504} (\bibinfo{year}{2001}).

\bibitem[13]{cheng-giordano-02}
\bibinfo{author}{\bibfnamefont{J.~T.} \bibnamefont{Cheng}} \bibnamefont{and}
  \bibinfo{author}{\bibfnamefont{N.}~\bibnamefont{Giordano}},
  \bibinfo{journal}{Phys. Rev. E} \textbf{\bibinfo{volume}{65}},
  \bibinfo{pages}{031206} (\bibinfo{year}{2002}).

\bibitem[14]{bib:nagayama-cheng-2004}
\bibinfo{author}{\bibfnamefont{G.}~\bibnamefont{Nagayama}} \bibnamefont{and}
  \bibinfo{author}{\bibfnamefont{P.}~\bibnamefont{Cheng}},
  \bibinfo{journal}{Int. J. Heat Mass Transfer} \textbf{\bibinfo{volume}{47}},
  \bibinfo{pages}{501} (\bibinfo{year}{2004}).

\bibitem[15]{bib:succi}
\bibinfo{author}{\bibfnamefont{S.}~\bibnamefont{Succi}},
  \emph{\bibinfo{title}{The Lattice {B}oltzmann Equation for Fluid Dynamics and
  Beyond}} (\bibinfo{publisher}{Oxford University Press},
  \bibinfo{year}{2001}).

\bibitem[16]{bib:nie-doolen-chen}
\bibinfo{author}{\bibfnamefont{X.}~\bibnamefont{Nie}},
  \bibinfo{author}{\bibfnamefont{G.~D.} \bibnamefont{Doolen}},
  \bibnamefont{and} \bibinfo{author}{\bibfnamefont{S.}~\bibnamefont{Chen}},
  \bibinfo{journal}{J. Stat. Phys.} \textbf{\bibinfo{volume}{107}},
  \bibinfo{pages}{279} (\bibinfo{year}{2002}).

\bibitem[17]{bib:shan-chen}
\bibinfo{author}{\bibfnamefont{X.}~\bibnamefont{Shan}} \bibnamefont{and}
  \bibinfo{author}{\bibfnamefont{H.}~\bibnamefont{Chen}},
  \bibinfo{journal}{Phys. Rev. E} \textbf{\bibinfo{volume}{47}},
  \bibinfo{pages}{1815} (\bibinfo{year}{1993}).
%
\bibinfo{author}{\bibfnamefont{X.}~\bibnamefont{Shan}} \bibnamefont{and}
  \bibinfo{author}{\bibfnamefont{H.}~\bibnamefont{Chen}},
  \bibinfo{journal}{Phys. Rev. E} \textbf{\bibinfo{volume}{49}},
  \bibinfo{pages}{2941} (\bibinfo{year}{1994}).

\bibitem[18]{bib:mplb}
\bibinfo{author}{\bibfnamefont{R.}~\bibnamefont{Benzi}},
  \bibinfo{author}{\bibfnamefont{S.}~\bibnamefont{Succi}}, \bibnamefont{and}
  \bibinfo{author}{\bibfnamefont{M.}~\bibnamefont{Vergassola}},
  \bibinfo{journal}{Phys. Rep.} \textbf{\bibinfo{volume}{222}},
  \bibinfo{pages}{145 } (\bibinfo{year}{1992}).
%
%
\bibinfo{author}{\bibfnamefont{S.}~\bibnamefont{Chen}} \bibnamefont{and}
  \bibinfo{author}{\bibfnamefont{G.}~\bibnamefont{Doolen}},
  \bibinfo{journal}{Ann. Rev. Fluid Mech.}
  \textbf{\bibinfo{volume}{30}}, \bibinfo{pages}{329} (\bibinfo{year}{1998}).

%
\bibitem[19]{bib:lbneq}
\bibinfo{author}{\bibfnamefont{S.}~\bibnamefont{Chen}},
  \bibinfo{author}{\bibfnamefont{H.}~\bibnamefont{Chen}},
  \bibinfo{author}{\bibfnamefont{D.}~\bibnamefont{Mart\'{\i}nez}},
  \bibnamefont{and}
  \bibinfo{author}{\bibfnamefont{W.}~\bibnamefont{Matthaeus}},
  \bibinfo{journal}{Phys. Rev. Lett.} \textbf{\bibinfo{volume}{67}},
  \bibinfo{pages}{3776} (\bibinfo{year}{1991}).
%
\bibinfo{author}{\bibfnamefont{H.}~\bibnamefont{Chen}},
  \bibinfo{author}{\bibfnamefont{S.}~\bibnamefont{Chen}}, \bibnamefont{and}
  \bibinfo{author}{\bibfnamefont{W.~H.} \bibnamefont{Matthaeus}},
  \bibinfo{journal}{Phys. Rev. A} \textbf{\bibinfo{volume}{45}},
  \bibinfo{pages}{R5339} (\bibinfo{year}{1992}).
%

%
\bibitem[20]{bib:contact-angle}
\bibinfo{author}{\bibfnamefont{J.}~\bibnamefont{Zhang}} \bibnamefont{and}
  \bibinfo{author}{\bibfnamefont{D.~Y.} \bibnamefont{Kwok}},
  \bibinfo{journal}{Phys. Rev. E} \textbf{\bibinfo{volume}{70}},
  \bibinfo{pages}{056701} (\bibinfo{year}{2004}).
%
\bibinfo{author}{\bibfnamefont{P.~G.} \bibnamefont{{De Gennes}}},
  \bibinfo{journal}{Rev. Mod. Phys.} \textbf{\bibinfo{volume}{57}},
  \bibinfo{pages}{827} (\bibinfo{year}{1985}).

\bibitem[21]{bib:benzi-06}
\bibinfo{author}{\bibfnamefont{R.}~\bibnamefont{Benzi}},
  \bibinfo{author}{\bibfnamefont{L.}~\bibnamefont{Biferale}},
  \bibinfo{author}{\bibfnamefont{M.}~\bibnamefont{Sbragaglia}},
  \bibinfo{author}{\bibfnamefont{S.}~\bibnamefont{Succi}},
\bibnamefont{and}
  \bibinfo{author}{\bibfnamefont{F.}~\bibnamefont{Toschi}},
  \bibinfo{journal}{arxiv:nlin.CD} \textbf{\bibinfo{volume}{nlin.CD}},
  \bibinfo{pages}{0602008} (\bibinfo{year}{2006}).

\bibitem[22]{benzi-biferale-05}
\bibinfo{author}{\bibfnamefont{R.}~\bibnamefont{Benzi}},
  \bibinfo{author}{\bibfnamefont{L.}~\bibnamefont{Biferale}},
  \bibinfo{author}{\bibfnamefont{M.}~\bibnamefont{Sbragaglia}},
  \bibinfo{author}{\bibfnamefont{S.}~\bibnamefont{Succi}},
\bibnamefont{and}
  \bibinfo{author}{\bibfnamefont{F.}~\bibnamefont{Toschi}},
  \bibinfo{journal}{Europhys. Lett.} \textbf{\bibinfo{volume}{73}},
  \bibinfo{pages}{in press} (\bibinfo{year}{2006}).

\bibitem[23]{bib:Navier}
\bibinfo{author}{\bibfnamefont{C.~L. M.~H.} \bibnamefont{Navier}},
  \bibinfo{journal}{Memoirs de l'Academie Royale des Sciences de l'Institut de
  France} \textbf{\bibinfo{volume}{1}}, \bibinfo{pages}{414}
  (\bibinfo{year}{1823}).

\bibitem[24]{cheikh-koper-03}
\bibinfo{author}{\bibfnamefont{C.}~\bibnamefont{Cheikh}} \bibnamefont{and}
  \bibinfo{author}{\bibfnamefont{G.}~\bibnamefont{Koper}},
  \bibinfo{journal}{Phys. Rev. Lett.} \textbf{\bibinfo{volume}{91}},
  \bibinfo{pages}{156102} (\bibinfo{year}{2003}).

\bibitem[25]{bib:cottin-bizone}
\bibinfo{author}{\bibfnamefont{C.}~\bibnamefont{Cottin-Bizonne}},
  \bibinfo{author}{\bibfnamefont{S.}~\bibnamefont{Jurine}},
  \bibinfo{author}{\bibfnamefont{J.}~\bibnamefont{Baudry}},
  \bibinfo{author}{\bibfnamefont{J.}~\bibnamefont{Crassous}},
  \bibinfo{author}{\bibfnamefont{F.}~\bibnamefont{Restagno}}, \bibnamefont{and}
  \bibinfo{author}{\bibfnamefont{E.}~\bibnamefont{Charlaix}},
  \bibinfo{journal}{Eur. Phys. J. E} \textbf{\bibinfo{volume}{9}},
  \bibinfo{pages}{47} (\bibinfo{year}{2002}).
%
\bibinfo{author}{\bibfnamefont{C.}~\bibnamefont{Cottin-Bizonne}},
  \bibinfo{author}{\bibfnamefont{C.}~\bibnamefont{Barentin}},
  \bibinfo{author}{\bibfnamefont{E.}~\bibnamefont{Charlaix}},
  \bibinfo{author}{\bibfnamefont{L.}~\bibnamefont{Bocquet}}, \bibnamefont{and}
  \bibinfo{author}{\bibfnamefont{J.}~\bibnamefont{Barrat}},
  \bibinfo{journal}{Eur. Phys. J. E} \textbf{\bibinfo{volume}{15}},
  \bibinfo{pages}{427} (\bibinfo{year}{2004}).

\bibitem[26]{baudry-charlaix-01}
\bibinfo{author}{\bibfnamefont{J.}~\bibnamefont{Baudry}} \bibnamefont{and}
  \bibinfo{author}{\bibfnamefont{E.}~\bibnamefont{Charlaix}},
  \bibinfo{journal}{Langmuir} \textbf{\bibinfo{volume}{17}},
  \bibinfo{pages}{5232} (\bibinfo{year}{2001}).


\bibitem[27]{toschi-succi-05}
\bibinfo{author}{\bibfnamefont{F.}~\bibnamefont{Toschi}} \bibnamefont{and}
  \bibinfo{author}{\bibfnamefont{S.}~\bibnamefont{Succi}},
  \bibinfo{journal}{Europhys. Lett.} \textbf{\bibinfo{volume}{69}},
  \bibinfo{pages}{549} (\bibinfo{year}{2005}).

\end{thebibliography}

\end{document}